\documentclass[letterpaper,twocolumn,english,aps,pra,floatfix,superscriptaddress]{revtex4}
\usepackage[T1]{fontenc}
\usepackage[latin9]{inputenc}
\usepackage{amsmath}
\usepackage{amssymb}
\usepackage{esint}

\makeatletter

\pdfpageheight\paperheight
\pdfpagewidth\paperwidth

\@ifundefined{textcolor}{}
{%
 \definecolor{BLACK}{gray}{0}
 \definecolor{WHITE}{gray}{1}
 \definecolor{RED}{rgb}{1,0,0}
 \definecolor{GREEN}{rgb}{0,1,0}
 \definecolor{BLUE}{rgb}{0,0,1}
 \definecolor{CYAN}{cmyk}{1,0,0,0}
 \definecolor{MAGENTA}{cmyk}{0,1,0,0}
 \definecolor{YELLOW}{cmyk}{0,0,1,0}
}

\renewcommand{\[}{\begin{equation}}
\renewcommand{\]}{\end{equation}}

\usepackage{babel}

\usepackage{babel}

\makeatother

\usepackage{babel}
\begin{document}
\global\long\def\avg#1{\langle#1\rangle}

\global\long\def\p{\prime}

\global\long\def\dg{\dagger}

\global\long\def\ket#1{|#1\rangle}

\global\long\def\bra#1{\langle#1|}

\global\long\def\proj#1#2{|#1\rangle\langle#2|}

\global\long\def\inner#1#2{\langle#1|#2\rangle}

\global\long\def\tr{\mathrm{tr}}

\global\long\def\pd#1#2{\frac{\partial#1}{\partial#2}}

\global\long\def\spd#1#2{\frac{\partial^{2}#1}{\partial#2^{2}}}

\global\long\def\der#1#2{\frac{d#1}{d#2}}

\global\long\def\im{\imath}

\global\long\def\S{\mathcal{S}}

\global\long\def\A{\mathcal{A}}

\global\long\def\F{\mathcal{F}}

\global\long\def\E{\mathcal{E}}

\global\long\def\As{{^{\sharp}}\hspace{-1mm}\mathcal{A}}

\global\long\def\Fs{{^{\sharp}}\hspace{-0.7mm}\mathcal{F}}

\global\long\def\Es{{^{\sharp}}\hspace{-0.5mm}\mathcal{E}}

\global\long\def\EsG{{^{\sharp}}\hspace{-0.5mm}\mathcal{E}_{G}}

\global\long\def\EsB{{^{\sharp}}\hspace{-0.5mm}\mathcal{E}_{B}}

\global\long\def\FsG{{^{\sharp}}\hspace{-0.5mm}\F_{G}}

\global\long\def\FsB{{^{\sharp}}\hspace{-0.5mm}\F_{B}}

\global\long\def\Fd{{^{\sharp}}\hspace{-0.7mm}\mathcal{F}_{\delta}}

\global\long\def\EG{\mathcal{E}_{G}}

\global\long\def\EB{\mathcal{E}_{B}}

\global\long\def\O{\mathcal{O}}

\global\long\def\SgF{\S d\F}

\global\long\def\SgEF{\S d\left(\E/\F\right)}

\global\long\def\U{\mathcal{U}}

\global\long\def\V{\mathcal{V}}

\global\long\def\H{\mathbf{H}}

\global\long\def\SO{\Pi_{\S}}

\global\long\def\PO{\hat{\Pi}_{\S}}

\global\long\def\SSH{\tilde{\Pi}_{\S}}

\global\long\def\EO{\Upsilon_{k}}

\global\long\def\ESH{\Omega_{k}}

\global\long\def\HSF{\mathbf{H}_{\S\F}}

\global\long\def\HSEF{\mathbf{H}_{\S\E/\F}}

\global\long\def\HS{\mathbf{H}_{\S}}

\global\long\def\ES{H_{\S}(t)}

\global\long\def\ESo{H_{\S}(0)}

\global\long\def\EgF{H_{\SgF} (t)}

\global\long\def\EgE{H_{\S d\E}(t)}

\global\long\def\EgEF{H_{\SgEF} (t)}

\global\long\def\EF{H_{\F}(t)}

\global\long\def\EFo{H_{\F}(0)}

\global\long\def\ESF{H_{\S\F}(t)}

\global\long\def\ESEF{H_{\S\E/\F}(t)}

\global\long\def\ESSEF{H_{\tilde{\S}\S\E/\F}(t)}

\global\long\def\EEFo{H_{\E/\F}(0)}

\global\long\def\EEF{H_{\E/\F}(t)}

\global\long\def\MI{I\left(\S:\F\right)}

\global\long\def\aMI{\left\langle \MI\right\rangle _{\Fs}}

\global\long\def\BS{\Pi_{\S} }

\global\long\def\PB{\hat{\Pi}_{\S} }

\global\long\def\QD{\mathcal{D}\left(\Pi_{\S}:\F\right)}

\global\long\def\QDp{\mathcal{D}\left(\PB:\F\right)}

\global\long\def\JI{J\left(\Pi_{\S}:\F\right)}

\global\long\def\CI{H\left(\F\left|\Pi_{\S}\right.\right)}

\global\long\def\CIp{H\left(\F\left|\PB\right.\right)}

\global\long\def\CS{\rho_{\F\left|s\right.}}

\global\long\def\CSu{\tilde{\rho}_{\F\left|s\right.}}

\global\long\def\CSp{\rho_{\F\left|\hat{s}\right.}}

\global\long\def\CEF{H_{\F\left|s\right.}}

\global\long\def\CEFp{H_{\F\left|\hat{s}\right.}}

\global\long\def\psiz{\ket{\psi_{\E\left|0\right.\hspace{-0.4mm}}}}

\global\long\def\psio{\ket{\psi_{\E\left|1\right.\hspace{-0.4mm}}}}

\global\long\def\psiinner{\inner{\psi_{\E\left|0\right.\hspace{-0.4mm}}}{\psi_{\E\left|1\right.\hspace{-0.4mm}}}}

\global\long\def\QDz{\boldsymbol{\delta}\left(\S:\F\right)_{\left\{  \sigma_{\S}^{z}\right\}  }}

\global\long\def\NQD{\bar{\boldsymbol{\delta}}\left(\S:\F\right)_{\BS}}

\global\long\def\EFS{H_{\F\left| \BS\right. }(t)}

\global\long\def\EFSM{H_{\F\left| \left\{  \ket m\right\}  \right. }(t)}

\global\long\def\Hol{\chi\left(\Pi_{\S}:\F\right)}

\global\long\def\Holp{\chi\left(\PB:\F\right)}

\global\long\def\ch{\raisebox{0.5ex}{\mbox{\ensuremath{\chi}}}_{\mathrm{Pointer}}}

\global\long\def\rhoS{\rho_{\S}(t)}

\global\long\def\rhoSo{\rho_{\S}(0)}

\global\long\def\rhoSF{\rho_{\S\F} (t)}

\global\long\def\rhoSgEF{\rho_{\SgEF} (t)}

\global\long\def\rhoSgF{\rho_{\SgF} (t)}

\global\long\def\rhoF{\rho_{\F}(t)}

\global\long\def\rhoFp{\rho_{\F}(\pi/2)}

\global\long\def\LE{\Lambda_{\E}(t)}

\global\long\def\LEc{\Lambda_{\E}^{\star}(t)}

\global\long\def\LEij{\Lambda_{\E}^{ij}(t)}

\global\long\def\LF{\Lambda_{\F}(t)}

\global\long\def\LFij{\Lambda_{\F}^{ij} (t)}

\global\long\def\LFc{\Lambda_{\F}^{\star}(t)}

\global\long\def\LEF{\Lambda_{\E/\F} (t)}

\global\long\def\LEFij{\Lambda_{\E/\F}^{ij}(t)}

\global\long\def\LEFc{\Lambda_{\E/\F}^{\star}(t)}

\global\long\def\Lkij{\Lambda_{k}^{ij}(t)}

\global\long\def\Hb{H}

\global\long\def\kE{\kappa_{\E}(t)}

\global\long\def\kEF{\kappa_{\E/\F}(t)}

\global\long\def\kF{\kappa_{\F}(t)}

\global\long\def\ts{t=\pi/2}

\global\long\def\QCB{\bar{\xi}_{QCB}}

\global\long\def\mc#1{\mathcal{#1}}

\global\long\def\MD{\lambda}

\global\long\def\up{\uparrow}

\global\long\def\down{\downarrow}

\global\long\def\Cku{\rho_{k\left|\up\right.}}

\global\long\def\Ckd{\rho_{k\left|\down\right.}}

\global\long\def\f{\mathcal{J}}

\global\long\def\onlinecite#1{\cite{#1}}

\title{Redundancy of einselected information in quantum Darwinism: The irrelevance
of irrelevant environment bits}

\author{Michael Zwolak}
\email{mpzwolak@gmail.com}

\address{Department of Physics, Oregon State University, Corvallis, OR 97331}

\author{Wojciech H. Zurek}

\address{Theoretical Division, Los Alamos National Laboratory, Los Alamos,
NM 87545}
\begin{abstract}
The objective, classical world emerges from the underlying quantum
substrate via the proliferation of redundant copies of selected information
into the environment, which acts as a communication channel, transmitting
that information to observers. These copies are independently accessible,
allowing many observers to reach consensus about the state of a quantum
system via its imprints in the environment. Quantum Darwinism recognizes
that the redundancy of information is thus central to the emergence
of objective reality in the quantum world. However, in addition to
the ``quantum system of interest,''
there are many other systems ``of no interest'' in the Universe
that can imprint information on the common environment. There is therefore
a danger that the information of interest will be diluted with irrelevant
bits, suppressing the redundancy responsible for objectivity. We show
that mixing of the relevant (the ``wheat'')
and irrelevant (the ``chaff'') bits
of information makes little quantitative difference to the redundancy
of the information of interest. Thus, we demonstrate that it does
not matter whether one separates the relevant information) from the
(irrelevant) chaff: The large redundancy of the relevant information
survives dilution, providing evidence of the objective, effectively
classical world.
\end{abstract}
\maketitle
Amplification \textendash{} already invoked by Bohr \cite{Bohr58-1}
\textendash{} is the central process by which the underlying quantum
substrate gives rise to the objective, classical world \cite{Zwolak14-1,Zwolak16-1}.
Quantum Darwinism \cite{Zurek09-1,Zurek14-1} formalizes this notion
into the concept of redundancy: When quantum systems are decohered
\cite{Joos03-1,Zurek03-1,Schlosshauer08-1}, they transfer select
information \textendash{} information about their pointer states \cite{Zurek81-1}
\textendash{} to their environment. Many observers can then infer
the state of the system indirectly by intercepting some small fragment
of the environment \cite{Ollivier04-1,Blume06-1}. In other words,
this select information is redundant, as any small fragment will do,
and is thus objective: Many observers can independently deduce the
pointer state of the system and reach consensus about it.

In our Universe there are many fragments of the environment that have
no or nearly no information about any given quantum system of interest
at any given time. Thus, in what way can one then apply the quantum
Darwinist considerations? To begin addressing this question, we consider
a spin model introduced in Ref. \cite{Blume05-1} with two types of
spins in the environment $\E$, ones that acquire perfect (classical)
information about the system $\S$ and others that acquire no information.
These are the ``good'' $\EG$ and ``bad'' $\EB$ environments,
respectively. This can be represented by a state of the form 
\begin{equation}
\left(\frac{1}{\sqrt{2}}\ket 0_{\S}\overbrace{\ket{0\cdots0}}^{\EG}+\frac{1}{\sqrt{2}}\ket 1_{\S}\overbrace{\ket{1\cdots1}}^{\EG}\right)\overbrace{\ket{0\cdots0}}^{\EB}.\label{eq:State}
\end{equation}
Physically, this state is generated when a set $\EG$ of ``good''
environment components \textendash{} the ``wheat'' \textendash{}
each perfectly decohere the system in the $z$-basis and a set $\EB$
\textendash{} the ``chaff'' \textendash{} do not interact at all
with the system. In what follows, the results are not limited to ``diametrically
opposed'' good and bad environments (i.e., ones like in Eq. \eqref{eq:State}),
but rather extend to the case with partial information in both the
good and bad environments, as well as mixed states, states without
permutational invariance (e.g., in the environment $\E_{G}$), and
alternative measurements to extract the information \cite{Kincaid17-1}.

Pure decoherence \textendash{} whether in a globally pure state or
in mixed state \textendash{} is the process by which the state structure
in Eq. \eqref{eq:State} arises. The Hamiltonian and initial states
for pure decoherence by independent environment components \cite{Zwolak09-1,Zwolak10-1}
are $\H=\H_{\S}+\PB\sum_{k=1}^{\Es}\EO+\sum_{k=1}^{\Es}\ESH$ with
$\left[\PB,\HS\right]=0$ and $\rho\left(0\right)=\rho_{\S}\left(0\right)\otimes\left[\bigotimes_{k=1}^{\Es}\rho_{k}\left(0\right)\right],$
where $k$ specifies an environment subsystem. Under pure decoherence,
no transitions occur between the pointer states $\hat{s}$ (the eigenstates
of $\PB$ \cite{Zurek03-1,Zurek81-1}). Up to unimportant local unitary
rotations, the state, Eq. \eqref{eq:State}, develops via a pure decoherence
process, e.g., one having a $g_{k}\sigma_{\S}^{z}\sigma_{k}^{z}$
interaction in the Hamiltonian with $g_{k}=1$ for $k\in\EG$ and
$g_{k}=0$ for $k\in\EB$, and an initial state $\left(\ket 0_{\S}+\ket 1_{\S}\right)\ket{+\cdots+}\ket{0\cdots0}$/2,
where $\ket +$ is a $\sigma^{x}$ eigenstate. These models \textendash{}
which include run-of-the-mill, everyday photon environments \cite{Zwolak14-1}
\textendash{} approximate the case where decoherence is strong compared
to the natural dynamics of the system. Moreover, spin models of this
type help elucidate the nature of redundancy in various settings \cite{Zwolak16-1,Riedel12-1},
which is what we will do here.

Intuitively, we know the redundancy of information in the state in
Eq. \eqref{eq:State}: There are $\EsG$ ``good'' bits, which are
in a GHZ state, and thus perfectly classically correlated with the
pointer observable of the system, and there are $\EsB$ ``bad''
bits, which are in a product state, and thus not correlated at all
with the system. Hence, the redundancy is just $\EsG$. However, in
a world where we are bombarded with good and bad bits alike, the question
arises: What is the typical fragment size that we need to intercept
from the total environment, $\E=\EG\otimes\EB$, to get nearly complete
information about $\S$? Likewise, in what way should we define redundancy,
with respect to $\E$ or just $\EG$? 

To answer these questions, we examine the mutual information, which
quantifies the correlations between the system and some fragment of
the environment, 
\[
\MI=H_{\S}+H_{\F}-H_{\S\F}.
\]
This can be divided into classical (the Holevo quantity \cite{Kholevo73-1})
and quantum (the discord \cite{Zurek00-1,Ollivier02-1,Henderson01-1})
components \cite{Zwolak13-1},
\[
\MI=\Hol+\QD,
\]
where $\BS$ specifies a basis (or, more generally, a POVM) on $\S$.
The Holevo quantity, $\Hol$, gives the maximum classical information
available about $\BS$ in $\F$, while the quantum discord $\QD$
is what remains. The information most efficiently transmitted by the
environment will be about the pointer basis, $\PB$ \cite{Ollivier04-1,Zwolak13-1}.
Thus, the Holevo quantity for that basis, $\Holp$, will be of interest.

To define redundant information, we need to know how much (what portion)
of the ``classical information of interest'' is contained in a typical
fragment $\F$ (of size $\Fd<\Es$) of the total environment, $\E$.
Thus, we seek $\Fd$, the size of the fragments that contain all but
the information deficit $\delta$ of the classical information, 
\begin{equation}
\avg{\Holp}_{\Fd}\simeq\left(1-\delta\right)H_{\S}.\label{eq:CondRed}
\end{equation}
Above $\avg{\cdot}_{\Fd}$ is the average over fragments of size $\Fd$
and $H_{\S}=H(\PB)$, i.e., the entropy of the pointer observable
\textendash{} the missing information about $\S$. Observers do not
require (and usually cannot get) all the missing information. The
information deficit, $\delta$, is the amount of information that
observers are prepared to forgo. The averaging can be done over the
relevant environment, $\EG$, or the total environment, $\E$. Alternatively,
one can maximize the number of distinct fragments that give nearly
complete information about $\PB$, i.e., $\left(1-\delta\right)H_{\S}$. 

Each of these approaches may yield different results. We will show
that all these procedures give the same value for the redundancy,
\begin{equation}
R_{\delta}=\frac{\Es}{\Fd},\label{eq:Red}
\end{equation}
up to an insignificant scaling factor, where $\Fd$ is taken from
Eq. \eqref{eq:CondRed} or from the maximization procedure. 

The Holevo quantity will approach $H_{\S}$ according to \cite{Zwolak14-1,Zwolak16-1}
\begin{equation}
\Holp\sim H_{\S}-H\left(P_{e}\right),\label{eq:Fano}
\end{equation}
where $P_{e}$, a function of $\F$, is the error probability for
distinguishing the conditional states of the fragment. The latter
are given by $\bra{\hat{s}}\rho_{\S\F}\ket{\hat{s}}/p_{\hat{s}}$,
where $\ket{\hat{s}}$ is a pointer state, $p_{\hat{s}}$ is the probability
that state occurs, and $\rho_{\S\F}$ is the reduced state of the
system and fragment. In the case of the state in Eq. \eqref{eq:State},
the conditional states are just $\ket{0\cdots0}_{\EG}\ket{0\cdots0}_{\EB}$
and $\ket{1\cdots1}_{\EG}\ket{0\cdots0}_{\EB}$.

The asymptotic behavior of the error probability is given by
\begin{equation}
P_{e}\sim\exp\left[-\QCB\Fs\right],\label{eq:Pe}
\end{equation}
where the exponent 
\begin{equation}
\QCB=-\ln\avg{\tr\left[\rho_{k\left|1\right.}^{c}\rho_{k\left|2\right.}^{1-c}\right]}_{k\in\E}\label{eq:TQCB}
\end{equation}
is the ``typical'' Chernoff information \cite{Zwolak14-1}, which
generalizes the quantum Chernoff bound (QCB) \cite{Audenaert07-1,Audenaert08-1,Nussbaum09-1,Li2016-1}
to sources of quantum states that are not independent and identically
distributed (i.i.d.). The error probability also depends on the $p_{\hat{s}}$,
but only in a prefactor to the exponential, and thus it does not play
a role as $\Fs$ becomes large. The optimal value of $c$ (with $0\le c\le1$)
is the one that maximizes $\QCB$. The latter is not an easy task
for non-i.i.d states. However, for spin systems undergoing pure decoherence
(of which the state in Eq. \eqref{eq:State} is an example) the value
of $c$ is $1/2$ \cite{Zwolak16-1} (the value of $c$ can also be
found for certain classes of photon environments \cite{Zwolak14-1}). 

This maps the understanding of classical information communicated
by $\F$ (and $\E$) into a problem of understanding the distinguishability
of conditional states, $\rho_{k\left|\hat{s}\right.}$ on individual
environment components $k$ (i.e., for the case here, individual environment
spins). Equations \eqref{eq:CondRed}, \eqref{eq:Fano}, and \eqref{eq:Pe}
allow us to estimate the redundancy \cite{Zwolak14-1,Zwolak16-1},
Eq. \eqref{eq:Red}, as 
\begin{equation}
R_{\delta}\simeq\Es\frac{\bar{\xi}_{QCB}}{\ln1/\delta}.\label{eq:QCBSupp}
\end{equation}
Indeed, this shows that macroscopic redundancy is unavoidable for
pure decoherence \textendash{} except for states of measure zero (i.e.,
completely mixed initial environment states or ones that commute with
the Hamiltonian), redundancy is always present \cite{Zwolak09-1,Zwolak10-1,Zwolak14-1,Zwolak16-1}.
The worst scenario for the dilution of information is when $\EsG$
is small and fixed, while $\EsB$ is taken to be larger and larger.
Taking $\EsB\gg\EsG,\,\Fs$, Eq. \eqref{eq:TQCB} is simple, 
\begin{align}
\QCB & =-\ln\left[\frac{\EsB\cdot1+\EsG\cdot0}{\Es}\right]\nonumber \\
 & =-\ln\left[\frac{\Es-\EsG}{\Es}\right]\simeq\frac{\EsG}{\Es}.\label{eq:QCBExpand}
\end{align}
Thus, the redundancy is 
\begin{equation}
R_{\delta}\simeq\frac{\EsG}{\ln1/\delta}.\label{eq:GBperf}
\end{equation}
We see that the calculation requires that $\EsG\ge\ln\left(1/\delta\right)$,
as redundancy can not be less than one. If $\EsG<\ln\left(1/\delta\right)$,
it means that the scenario is outside of the realm of validity of
the QCB calculation, as an observer needs essentially the whole environment
to approach $\left(1-\delta\right)H_{\S}$ bits of information about
$\S$, if they can acquire that amount of information at all. 

We note that one can also \emph{exactly} solve for the average Holevo
quantity or mutual information, which yields the same result as above:
Given a fragment $\F$, $\Holp$ will be one if a good bit is intercepted
and zero otherwise (similarly for the quantum mutual information,
$\MI$, unless all the good bits are intercepted, but this happens
with negligible probability). The probability that the observer will
intercept $\Fs$ bad bits \textendash{} and have zero information
about the system \textendash{} is 
\[
P_{B}=\frac{\EsB}{\Es}\cdot\frac{\EsB-1}{\Es-1}\cdots\frac{\EsB-(\Fs-1)}{\Es-(\Fs-1)},
\]
with $\Es=\EsG+\EsB$. The probability to intercept at least one good
bit is $1-P_{B}$, giving 
\[
\left\langle \Holp\right\rangle _{\Fs}=H_{\S}\left(1-P_{B}\right)=1-\frac{\EsB!\left(\Es-\Fs\right)!}{\left(\EsB-\Fs\right)!\Es!},
\]
with $H_{\S}=1$. Redundancy requires that 
\begin{equation}
\delta\simeq\frac{\EsB!\left(\Es-\Fd\right)!}{\left(\EsB-\Fd\right)!\Es!}\label{eq:ExactDel}
\end{equation}
and we can use Stirling's approximation to get the limiting forms
of this expression, $\ln\delta\simeq-\Fd\EsG/\EsB.$ This yields the
redundancy 
\[
R_{\delta}=\frac{\Es}{\Fd}\simeq\frac{\EsB+\EsG}{\frac{\EsB}{\EsG}\ln1/\delta}\simeq\frac{\EsG}{\ln1/\delta},
\]
in agreement with the QCB result. For small $\Fd$, one can similarly
expand Eq. \eqref{eq:ExactDel}, finding $R_{\delta}=\Es_{G}/(1-\delta)\approx\EsG/\ln\left(1/\delta\right)$
for $\delta$ near 1 \textendash{} e.g., for $\Fd=1$, $\delta=\Es_{B}/\Es\approx1$
\textendash{} and showing that the QCB result can work well even in
the non-asymptotic regime. While this exact calculation is simple,
the QCB calculation is even simpler still and is easily extended to
many other cases (see, e.g., Ref. \cite{Zwolak16-1}). 

If instead of the usual definition of redundancy (Eqs. \eqref{eq:CondRed}
and \eqref{eq:Red}), one defined redundancy as the maximum number
of disjoint fragments for which $\Holp\simeq\left(1-\delta\right)H_{\S}$,
the result would be $R_{\delta}=\EsG$. For this ``perfect'' good-bad
state, Eq. \eqref{eq:State}, this is true for any $\delta\neq1$
\footnote{When $\delta=1$, $R_{\delta}=\Es$, as every component of the environment
contains at least \emph{zero} information.}. By the same token, if one averaged Eq. \eqref{eq:CondRed} only
over the environment $\EG$, then $\Fd=1$ and $R_{\delta}=\EsG$:
Only one spin from $\EG$ is necessary to acquire the requisite information
(we note that for the perfect GHZ state of the good bits with the
system, the QCB calculation is not valid when limiting to the relevant
environment, as $\QCB$ diverges, reflecting that $R_{\delta}=\EsG$
for any $\delta$). Indeed, these two latter approaches agree with
our intuition about the state in Eq. \eqref{eq:State}, as there are
just $\EsG$ copies of the pointer state information. 

These definitions lead to $R_{\delta}=\EsG/\ln\left(1/\delta\right)$
and $R_{\delta}=\Es_{G}$ and are thus not equivalent. However, they
only differ by a factor of $\ln1/\delta$ (and this result is unaffected
by the choice of $p_{0}$, $p_{1}$, the probabilities for $\ket 0_{\S}$
and $\ket 1_{\S}$ in the initial state of the system) \footnote{We note that Eq. \eqref{eq:QCBSupp} is an asymptotic lower bound
to the redundancy, but it was proven in Ref. \cite{Zwolak14-1} that
it is within a factor of 2 of the exact result. For all cases where
exact results can be found, such as the case here, Eq. \eqref{eq:QCBSupp}
is in agreement and is thus not an estimate. However, since the redundancy
computed from the maximization procedure is larger than Eq. \eqref{eq:QCBSupp},
this means that we are finding a worst case difference. Similarly,
we expanded the logarithm in Eq. \eqref{eq:QCBExpand}. This approximation
decreases the value of the redundancy. Without it, the difference
between the definitions of redundancy would be closer.}. For any reasonable $\delta$, the two definitions are practically
the same \footnote{For $\delta=\exp\left(-X\right)$, the definitions differ only by
a factor of $X$. Thus, for a very small $\delta$, e.g., $\delta=\exp\left(-10\right)$
or $\exp\left(-20\right)$, Eq. \eqref{eq:CondRed} underestimates
the redundancy by only a factor of, e.g., 10 or 20. In the opposite
regime, when $\delta$ is close, but not equal, to 1 (i.e., requiring
that the records hold very little information), the redundancy using
Eq. \eqref{eq:CondRed} continuously approaches $R_{\delta}=\Es$
as $\Fd\to1$. The maximization, though, changes discontinuously,
jumping from $R_{\delta}=\Es_{G}$ to $R_{\delta}=\Es$ at $\delta=1$
(where the records hold no information, which is an unimportant case).
When one is interested in $\delta$ being close (but not equal) to
one, using the maximization is more appropriate, as it accurately
reflects the number of imperfect copies of information, whereas using
Eq. \eqref{eq:CondRed} will overestimate this redundancy.}. The reason for the correspondence between the definitions is that \textendash{}
when an observer intercepts both relevant and irrelevant bits \textendash{}
the probability not to get any good bit drops exponentially with the
size of the fragment $\Fs$. This only weakly affects the ability
of the observer to capture a good bit and deduce the state of the
system. This is remarkable, as it says that when states of the form
in Eq. \eqref{eq:State} arise we need not worry about distinguishing
between parts of the larger environment \textendash{} parts that interact
with the system and parts that either do not or only weakly interact
\textendash{} for quantifying the redundancy of information. In other
words, we need not worry about separating the wheat from the chaff. 

This example can be extended to the case where the good and bad spins
are not diametrically opposed, i.e., not perfectly good or bad. For
instance, one can consider $\EsG$ good spins that contribute $\left|\gamma_{G}\right|^{2}$
to the decoherence factor and $\EsB$ bad spins that contribute $\left|\gamma_{B}\right|^{2}$.
In this case, the QCB gives immediately 
\begin{equation}
R_{\delta}\simeq\frac{\Es\ln\left[\frac{\EsB}{\Es}\left|\gamma_{B}\right|^{2}+\frac{\EsG}{\Es}\left|\gamma_{G}\right|^{2}\right]}{\ln\delta},\label{eq:QCBFullGoodBad}
\end{equation}
where we have made use of the relationship between decoherence and
information in pure states, $\tr\rho_{k\left|1\right.}^{c}\rho_{k\left|2\right.}^{1-c}=\left|\gamma_{k}\right|^{2}$
\cite{Zwolak16-1}. 

As with the ``perfect'' good-bad state, this calculation can be
done in an alternative manner. To find the averaged Holevo quantity,
one makes use of the equality $\Holp=H\left(\frac{1+\left|\gamma_{G}\right|^{\Fs-\FsB}\left|\gamma_{B}\right|^{\FsB}}{2}\right)$
for pure states or its more general form for $p_{0}\neq p_{1}$ \cite{Zwolak10-1}.
Expanding the binary entropy $H\left(x\right)$ for $x$ near 1/2
shows that one just needs to find the average decoherence factor.
For fragments of size $\Fs$, the latter is given by the sum of $\FsB$
from 0 to $\Fs$ of $\left|\gamma_{G}\right|^{2\left(\Fs-\FsB\right)}\left|\gamma_{B}\right|^{2\FsB}$
times the probability

\[
\frac{\Fs!}{\FsB!\left(\Fs-\FsB\right)}\frac{\EsB!}{\left(\EsB-\FsB\right)!}\frac{\EsG!}{\left(\EsG-\FsG\right)!}\frac{\left(\Es-\Fs\right)!}{\Es!}
\]
of intercepting $\FsB$ bad spins and $\FsG$ good spins in the fragment.
Stirling's approximation can be used on the three last factors in
the probability and the sum performed. This yields the QCB result,
Eq. \eqref{eq:QCBFullGoodBad}, at much greater expense. 

For simplicity, we now assume that the bad spins give $\left|\gamma_{B}\right|^{2}=1$,
i.e., a worst case where those bits never interacted with the system
(and thus have no information). Expanding the QCB result, Eq. \eqref{eq:QCBFullGoodBad},
one finds, 
\begin{equation}
R_{\delta}\simeq\frac{\EsG\left(1-\left|\gamma_{G}\right|^{2}\right)}{\ln1/\delta}.\label{eq:GoodBadRed}
\end{equation}

Maximizing the number of fragments that give $\left(1-\delta\right)H_{\S}$
bits of information is asymptotically equivalent to a QCB calculation
(for permutationally invariant states) that simply ignores the $\EsB$
bad spins: 
\begin{equation}
R_{\delta}\simeq\frac{\EsG\ln\left|\gamma_{G}\right|^{2}}{\ln\delta}.\label{eq:MaxRed}
\end{equation}
These results, Eqs. \eqref{eq:GoodBadRed} and \eqref{eq:MaxRed},
for the redundancy look different and it is not readily apparent that
the same conclusions hold as in the ``perfect'' good-bad model,
as the difference depends on the contribution to the decoherence factor
from the good spins. The ratio between the two results, Eqs. \eqref{eq:GoodBadRed}
and \eqref{eq:MaxRed}, is 
\begin{equation}
\frac{\left(1-\left|\gamma_{G}\right|^{2}\right)}{\ln1/\left|\gamma_{G}\right|^{2}}.\label{eq:Ratio}
\end{equation}
As we will now show, the smallest this ratio can be is $\sim1/\ln1/\delta$,
just like the perfect good-bad model above. 

One first has to note that Eq. \eqref{eq:Ratio} is monotonically
increasing to 1 as $\left|\gamma_{G}\right|^{2}$ increases to 1 (giving
a redundancy of zero and making the definitions of redundancy the
same in this limit). This can be proven by taking its derivative and
applying the inequality in footnote {[}25{]} of Ref. \cite{Zwolak14-1}
to show the derivate is always positive. The smallest value of Eq.
\eqref{eq:Ratio} can then be found by taking the smallest value of
$\left|\gamma_{G}\right|^{2}$ allowed by a consistent application
of the QCB. There are two calculations of redundancy \textendash{}
one with good and bad spins, and one with only good spins. The former,
Eq. \eqref{eq:GoodBadRed}, allows for all $\left|\gamma_{G}\right|^{2}$
(indeed, this just recovers Eq. \eqref{eq:GBperf}). The latter, Eq.
\eqref{eq:MaxRed}, however, requires $\delta\leq\left|\gamma_{G}\right|^{2}$
(i.e., Eq. \eqref{eq:MaxRed} can not be larger than $\EsG$ no matter
how small $\left|\gamma_{G}\right|^{2}$ is, or, in other words, $\Fd$
can not be less than 1). When 
\[
\frac{\ln\left|\gamma_{G}\right|^{2}}{\ln\delta}=1,
\]
or $\left|\gamma_{G}\right|^{2}=\delta$, each good spin holds a sufficient
record to immediately put the information to within $\delta$ of the
plateau upon receipt of one spin \footnote{This ignores finite size effects, which are nevertheless order 1.}.
When $\left|\gamma_{G}\right|^{2}<\delta$, the redundancy no longer
depends on $\left|\gamma_{G}\right|^{2}$, it is just $R_{\delta}=\EsG$
\footnote{In this regime, $\left|\gamma_{G}\right|^{2}<\delta$, one has to
perform computations in an alternative way, but the conclusion is
the same: The redundancy using the maximum is $R_{\delta}=\EsG$ and
the redundancy from the other definition does not change (the presence
of $\EB$ ``regularizes'' the QCB, so it does not matter if $\left|\gamma_{G}\right|^{2}<\delta$).
Thus, their minimum ratio is still Eq. \eqref{eq:MinRat}.}. Thus, the QCB result, Eq. \eqref{eq:MaxRed}, is valid only when
$\left|\gamma_{G}\right|^{2}\ge\delta$. When this is the case, the
ratio of the two computed redundancies is Eq. \eqref{eq:Ratio}. This
ratio is minimal when $\left|\gamma_{G}\right|^{2}=\delta$, giving
\begin{equation}
\frac{\left(1-\delta\right)}{\ln1/\delta}.\label{eq:MinRat}
\end{equation}
and is also unaffected by the choice of $p_{0}$ ($p_{1}$). Since
we want $\delta$ to be small, this ratio is essentially $1/\ln1/\delta$.
Thus, just as with the ``perfect'' good-bad model, the two definitions
differ only by an insignificant factor. 

In both cases, we see that \textendash{} when examining symmetric
(permutationally invariant) ``good'' environments, $\EG$ \textendash{}
there is no difference between redundancy defined as the maximization
and redundancy defined with the averaging in Eq. \eqref{eq:CondRed}
over only the good environment. To extend this calculation to mixed
and/or non-permutationally invariant states is a straightforward matter.
One only has to note the Chernoff Information is no longer directly
related to the decoherence factor, $\left|\gamma_{G}\right|^{2}$,
but rather takes on a different form (compare Eq. (16) and Eq. (19)
in Ref. \cite{Zwolak16-1}), but otherwise the calculation is formally
identical. Thus, the correspondence between definitions of redundancy
\textendash{} i.e., a difference of at most $\sim\ln1/\delta$ \textendash{}
is a general feature of pure decoherence. 

This example not only shows the ease of computation using the QCB,
it also helps us understand the definition of redundancy itself. If
redundancy was defined by maximizing the number of copies of information,
rather than taking an average over all fragments of a given size,
then one would obtain a different value. However, both definitions
give $R_{\delta}\propto\EsG$, where the proportionality is different
only by a factor of $\ln\left(1/\delta\right)$ \textendash{} i.e.,
a factor that is only weakly dependent on $\delta$. This is the case
even when taking an essentially arbitrarily large number of bad bits
in the environment. The reason for such a close correspondence between
definitions is that, as the observer intercepts a larger and larger
fragment, the probability of not receiving a good bit decreases exponentially
with fragment size. The observer is thus likely to always to receive
a good bit. The definition of redundancy in Eqs. \eqref{eq:CondRed}
and \eqref{eq:Red} applied to the total environment $\E$, therefore,
gives reasonable estimates \textendash{} lower than the maximal redundancy
but different only by an insignificant factor \textendash{} for the
number of records proliferated into the environment. Thus, the emergence
of the classical, objective reality is unavoidable \cite{Ollivier04-1,Zwolak14-1,Brandao15-1}
and there is no need to separate the wheat from the chaff to perceive
objective states of the systems of interest. 
\begin{acknowledgments}
We would like to thank Marek Rams for helpful discussions and the
Center for Integrated Quantum Science and Technology (IQST) and the
University of Ulm, where part of this work was carried out. This research
was supported in part by the U.S. Department of Energy through the
LANL/LDRD Program and by the Foundational Questions Institute Grant
No. 2015-144057 on \textquotedblleft Physics of What Happens.\textquotedblright{} 
\end{acknowledgments}

\end{document}